\documentclass[prd,nofootinbib,letterpaper,floatfix,superscriptaddress,preprintnumbers,twocolumn]{revtex4}

\usepackage{color}

\usepackage{graphicx}

\usepackage{verbatim}
\usepackage{amsmath}
\usepackage{amssymb}
\usepackage{subfigure}
\usepackage{url}

\usepackage{multirow}
\usepackage{threeparttable}
\usepackage{paralist}

\usepackage{color}
\definecolor{darkgreen}{rgb}{0,0.5,0}

\DeclareRobustCommand{\Sec}[1]{Sec.~\ref{#1}}

\DeclareRobustCommand{\Fig}[1]{Fig.~\ref{#1}}

\DeclareRobustCommand{\Eq}[1]{Eq.~(\ref{#1})}

\newcommand{\be}{\begin{equation}}
\newcommand{\ee}{\end{equation}}

\begin{document}

\title{A New Probe of Naturalness}
 
\author{Nathaniel Craig} \email{ncraig@ias.edu}
\affiliation{Institute for Advanced Study, Princeton, NJ 08540, USA \\
  Department of Physics and Astronomy, Rutgers University, Piscataway,
  NJ 08854, USA}

\author{Christoph Englert} \email{christoph.englert@durham.ac.uk}
\affiliation{Institute for Particle Physics Phenomenology, Department
  of Physics,\\
  Durham University, Durham DH1 3LE, UK}

\author{Matthew McCullough} \email{mccull@mit.edu} \affiliation{Center
  for Theoretical Physics, Massachusetts Institute of Technology,
  Cambridge, MA 02139, USA}

\date{\today}

\begin{abstract}
Any new scalar fields that perturbatively solve the hierarchy problem by stabilizing the Higgs mass also generate new contributions to the Higgs field-strength renormalization, irrespective of their gauge representation.  These new contributions are physical and their magnitude can be inferred from the requirement of quadratic divergence cancellation, hence they are directly related to the resolution of the hierarchy problem.  Upon canonically normalizing the Higgs field these new contributions lead to modifications of Higgs couplings which are typically great enough that the hierarchy problem and the concept of electroweak naturalness can be probed thoroughly within a precision Higgs program.  Specifically, at a Linear Collider this can be achieved through precision measurements of the Higgs associated production cross-section.  This would lead to indirect constraints on perturbative solutions to the hierarchy problem in the broadest sense, even if the relevant new fields are gauge singlets.

\end{abstract}

\preprint{IPPP/13/30, DCPT/13/60, MIT-CTP {4462}, RU-NHETC-{2013-12}}

\maketitle

\section{Introduction}
\label{sec:introduction}

The discovery of the Higgs at the LHC
\cite{ATLAS-CONF-2013-034,CMS-PAS-HIG-13-005} and lack of evidence for
physics beyond the Standard Model have heightened the
urgency of the electroweak hierarchy
problem.  This motivates focusing experimental searches towards testing
``naturalness from the bottom up'' as broadly as possible.
In practice this means generalizing beyond the specifics of particular
UV-complete models and instead constraining the additional degrees of
freedom whose couplings to the Higgs are responsible for canceling the
most pressing quadratically divergent Standard Model contributions to
the Higgs mass. While these couplings may appear tuned from the
perspective of the low-energy effective theory, we may assume they are
dictated by symmetries of the full theory. To a certain extent, this
strategy is already being pursued in searches for stops in SUSY and
$t'$ fermions, however the Standard Model gauge representations of top
partners are not necessarily fixed by the cancellation of quadratic
divergences. For example,
in twin Higgs models \cite{Chacko:2005pe} the degrees of freedom
protecting the Higgs mass are completely neutral under the Standard
Model, while in folded supersymmetry \cite{Burdman:2006tz} the scalar
top partners are neutral under QCD and only carry electroweak quantum
numbers. Such models provide proof of principle that the Higgs mass
may be protected by degrees of freedom that carry a variety of
Standard Model gauge charges, and there are likely to be broad classes
of theories with similar properties.

As we will discuss further in \Sec{sec:effective}, direct searches for
these additional degrees of freedom can be particularly challenging
depending on the gauge charges.  Therefore in this work we will
advocate an additional and complementary approach, concerned with
exploring naturalness \emph{indirectly}.  In certain cases this may be
the most promising avenue for constraining additional degrees of
freedom associated with the naturalness of the Higgs
potential.\footnote{For recent work probing naturalness indirectly
  when new fields are charged under QCD and
  contribute directly to Higgs digluon and Higgs diphoton couplings at one loop, see
  e.g. \cite{Arvanitaki:2011ck, Blum:2012ii, Montull:2012ik}.}

Specifically, we establish for the first time a quantitative connection 
between quadratically divergent Higgs mass corrections and 
new contributions to the Higgs wave-function renormalization in natural theories.  The latter are physical and 
modify Higgs couplings.

To illustrate the possible indirect effects of natural new
physics, consider a scenario where the Higgs is coupled to some new
top-partner fields that cancel the one-loop quadratic
divergences arising from top-quark loops.  \Eq{eq:corrections}
schematically indicates that, as well as the usual Higgs mass
corrections, one will also in general have corrections to the Higgs
wave-function renormalization\footnote{There are also typically
  corrections to the cubic and quartic couplings as well, which we do
  not show in this diagram.}
\begin{equation}
  \delta Z_h, \delta m_h^2~\sim~
  \parbox{3.3cm}{\vspace{-0.1cm}
    \includegraphics[width=3.3cm]{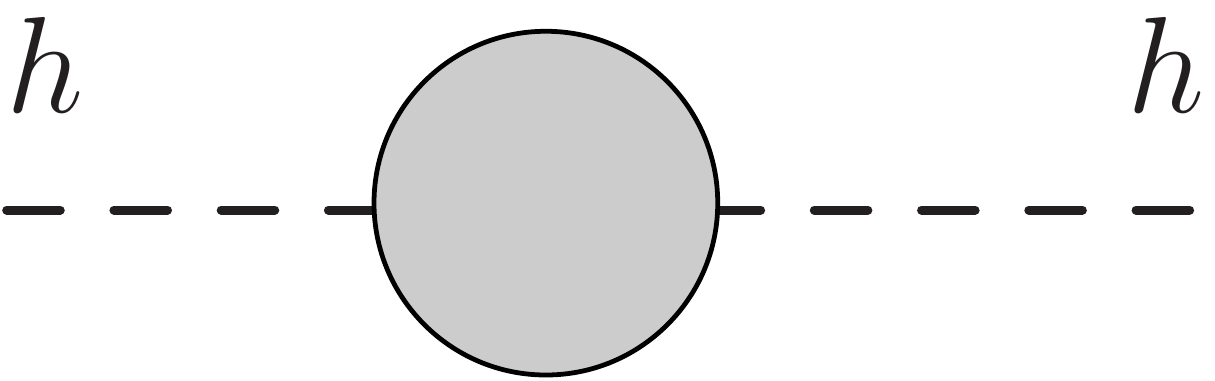}}~.
    \label{eq:corrections}
\end{equation}

At the Higgs mass-scale we may write the
full one-loop effective Lagrangian as
\be 
\label{eq:Z}
\mathcal{L} = \mathcal{L}_{SM} + \frac{1}{2} 
\delta Z_h (\partial_\mu h)^2 +...
\ee
where $\delta Z_h$ is directly related to the new quadratic Higgs mass corrections, 
$ \mathcal{L}_{SM}$ is the full SM Lagrangian at one loop, and
the ellipsis denote corrections to the Higgs mass, cubic and quartic
couplings coming from the new fields.\footnote{We have, for now, assumed that
the new fields are gauge singlets and so expect no corrections to
the weak fields or couplings other than Higgs
self-couplings.  We will discuss scenarios with
  non-gauge-singlet fields shortly.}  As with the precision
electroweak program
\cite{Peskin:1990zt,Peskin:1991sw,Altarelli:1990zd,Altarelli:1991fk,Kennedy:1988sn,Kennedy:1988rt}
we need to determine which corrections are physical and can be
constrained by measurement, and which are unphysical.  We first
canonically normalize the Higgs field by the re-scaling $h\rightarrow
(1-\delta Z_h/2) h$.  This re-scales all Higgs couplings and the mass
operator.  The Higgs cubic and quartic couplings have not been
measured directly, and so the new re-scaled values are unconstrained.
Also, the Higgs mass is a free parameter of the theory which can absorb this field re-definition. However all Higgs
couplings to weak gauge bosons and fermions have been re-scaled by the
same amount.  This re-scaling is physical: it can be moved around by
re-scaling other fields or couplings but cannot be removed from the
theory.  For canonically normalized fields this re-scaling will in
general break the SM prediction for the relationship between the mass
of a field and its coupling to the Higgs.  This deviation from SM predictions 
can then be constrained with precision Higgs coupling measurements.

In the case where the new fields are not gauge singlets one
expects additional corrections beyond the wave-function renormalization.  Some of these corrections involve the
gauge sector alone, and can be constrained via the Peskin-Takeuchi
parameters \cite{Peskin:1990zt,Peskin:1991sw} and their generalization
\cite{Barbieri:2004qk}; other corrections may also directly correct
the Higgs-weak boson vertices.  Although this situation is more
involved, the wave-function renormalization typically dominates
\cite{Englert:2013tya}.  Hence we see that if the hierarchy problem is
resolved by new physics then it may leave its footprint
through indirect signatures in SM processes via modified Higgs couplings, even in situations where
it is difficult to observe the new physics directly.

Thus far the discussion has been rather general.  To render these
effects quantitative, we must commit to a concrete, calculable set-up.
In \Sec{sec:effective} we will construct a general scenario based
solely on the naturalness criterion: a ``weak-scale effective theory
of naturalness,'' restricted only by the simplifying assumption that
the new fields canceling the top quadratic divergence are
scalars.\footnote{We note that a generalization to spin-1/2 or even
  spin-1 partners is also in principle possible.}  In \Sec{sec:LC} we
describe how, guided by \emph{naturalness alone}, one is led to very
specific quantitative predictions for Higgs coupling corrections within this effective natural
theory, with the only free variables being the number of fields and
their masses.  We will clearly demonstrate that, even if direct
evidence for a natural weak scale remains elusive, the generic
parameter space of natural theories can be thoroughly explored through
percent-level precision Higgs coupling measurements at a Linear
Collider (LC) or potentially at the LHC.

\section{Weak-Scale Effective Theory of Naturalness}
\label{sec:effective}
Assuming that the leading natural degrees of freedom are scalar top
partners we can define the perturbative effective natural theory as
\begin{equation} \label{eq:L}
  {\cal{L}}= {\cal{L}}_{\text{SM}} + 
  \sum_i \left( |\partial_\mu\phi_i|^2-m_i^2 |\phi_i|^2 
    -\lambda_i |H|^2 |\phi_i|^2 \right) ~~,
\end{equation}
where without loss of generality we take the scalars to be complex,
and we use the EW symmetry breaking conventions $H \to v + h
/\sqrt{2}$ with $v \approx 174$ GeV and $m_{\phi_i}^2 = m_i^2 +
\lambda_i v^2$, leading to a trilinear coupling ${\cal{L}} \supset
\sqrt{2} \lambda_i v h |\phi_i|^2$.\footnote{If the top partners are
  in weak doublets we could also have couplings such as $V \supset |H
  \cdot \phi|^2$, as in the MSSM for the left-handed top squark.
  However, since we are only really concerned with the couplings
  between top-partners and the neutral Higgs, \Eq{eq:L} still captures
  the relevant phenomenology.} Here the index $i = 1,\dots,n_\phi$
counts the number of fields coupled to $H$, which may be related by
gauge or global symmetries. For example, in SUSY $n_\phi = 6$ counts
the two top squarks transforming as triplets under $SU(3)_c$, while in
folded SUSY $n_\phi = 6$ counts the two top squirks transforming as
triplets under a distinct $SU(3)$ gauge group.

In order to cancel one-loop quadratic Higgs mass corrections from the
top quark alone it is simply required that 
\be 
\sum_i \lambda_i = 6
\lambda_t^2 ~~, 
\ee
where $\lambda_t$ is the top Yukawa coupling.\footnote{We will not be
  concerned with one-loop quadratic divergences from loops of gauge
  degrees of freedom, however if desired these loops could be
  cancelled by extra fermions, as in SUSY, or even by choosing a
  modified value of $\lambda_\phi$.}  For simplicity we can make the
further assumption that all $n_\phi$ scalars have the same mass,
$m_\phi$, and the same coupling $\lambda_\phi$.  As we will show, this
extremely simple effective theory of naturalness is broad enough to
capture the dominant indirect corrections to Higgs physics even though
we have not specified the gauge representations and are agnostic as to
the UV-completion of the model.

From this point we can define a measure of naturalness.  Although the
theory so far is renormalizable we should choose an energy scale,
$\Lambda$, at which the theory is UV-completed.\footnote{For example,
  in SUSY theories this would typically correspond to the
  SUSY-breaking messenger scale.}  We can then calculate corrections
to the high-scale Higgs mass, $m_H$, due to logarithmic running from
$\Lambda$ down to the weak scale.  At one loop this correction is
\begin{eqnarray}
  \delta m_H^2 & = & -n_\phi  \frac{\lambda_\phi}{8 \pi^2} m_\phi^2 \log \left( \frac{\Lambda}{m_\phi} \right) , \\
  & = & -\frac{6 \lambda_t^2}{8 \pi^2} m_\phi^2 \log \left( \frac{\Lambda}{m_\phi} \right) ,
\end{eqnarray}
where we have imposed the cancellation of quadratic divergences in the
final line and $m_\phi$ is the top partner mass above the EW-breaking scale.  Following standard conventions used in SUSY literature we
can define the fine-tuning measure as a function of the logarithmic corrections to the high-scale Higgs mass, $m_H$, and the physical Higgs mass, $m_h$, as \cite{Kitano:2006gv}
\be 
\Delta \equiv \frac{2
  \delta m_H^2}{m_h^2} ~~, 
\ee 
which quantifies the degree of fine-tuning required in the Higgs
potential.  Since we do not know the details of the UV-completion it
is sensible to assume a low UV-completion scale, so we set $\Lambda =
10$ TeV, although we note that in concrete models this scale can in
principle be much higher, exacerbating the fine-tuning.  With this
measure we can consider some benchmark tuning points
\begin{eqnarray}
\Delta^{-1} (m_\phi = 350 \text{ GeV}) & = & 25 \% \\
\Delta^{-1} (m_\phi = 605 \text{ GeV}) & = & 10 \%
\label{eq:benchmarks}
\end{eqnarray}
These benchmarks illustrate that for a natural weak scale scalar top
partners should not lie too far from the weak scale, regardless of the
particular UV completion or top partner gauge charges.

Although irrelevant to the one-loop corrections to the Higgs mass, the
fields $\phi$ may be charged under various representations of the
Standard Model gauge group. Some of these representations may be
searched for directly at the LHC. For example, states charged under
$SU(3)_c$ are primarily produced via QCD interactions and are
efficiently constrained by LHC stop searches or, if stable, searches
for $R$-hadrons. Similarly, states charged under $SU(2)_L$ and/or
$U(1)_Y$ are primarily produced via Drell-Yan processes and are
constrained by LHC searches for electroweak final states or, if
stable, CHAMP searches.

States neutral under the Standard Model are much more challenging to
constrain through direct searches. Although $\phi$ appears in the
invisible decay products of the Higgs when $m_{\phi} < m_h / 2$, the
coupling $\lambda_\phi$ is typically large enough that this invisible
partial width vastly exceeds the Standard Model width and is ruled out
by current limits on the invisible width of the Higgs
\cite{Giardino:2013bma} unless $m_{\phi}$ is finely tuned to lie at
the kinematic threshold for pair production. For $m_{\phi} > m_h / 2$,
the primary means of observing $\phi$ at the LHC involves pair
production through an off-shell Higgs boson. The most promising
channels are vector boson fusion, $q q \to V^\star V^\star qq \to \phi
\phi^* qq$, and vector associated production, $q \bar q \to V^\star
\to V \phi \phi^*$, where $V = W,Z$. However, the small production
cross-sections and the challenges of triggering and pileup for the
relevant final states render these direct search channels unpromising
at the LHC. Finally, although the lightest $\phi_i$ could constitute a
dark matter candidate if absolutely stable and neutral under the SM,
its thermal relic abundance is typically too small if governed by 
$s$-channel annihilation through the Higgs \cite{Djouadi:2011aa} due 
to the large coupling to the Higgs.  If this issue is circumvented via 
a non-thermal production mechanism and the top partner saturates the 
observed DM abundance then direct detection constraints rule out such 
large couplings \cite{Djouadi:2011aa}.

Note that even states carrying Standard Model gauge charges are
exceptionally difficult to discover if the kinematics of their decays are
unfavorable. Colored scalars decaying to nearly-degenerate neutral
states are challenging to distinguish from Standard Model di-jet
backgrounds. Electroweak scalars whose mass is close to the $W$ boson
are difficult to discover underneath $W^+ W^-$ backgrounds, while
decays to nearly-degenerate neutral states are challenging for
standard triggers.

Thus it is entirely possible that the mass of the Higgs boson is
rendered completely natural by top partners whose kinematic properties
or quantum numbers make them difficult to discover at the LHC and
perhaps the most promising avenue for discovery lies in {\it indirect}
searches.

\section{A New Probe of Naturalness}
\label{sec:LC}
An efficient indirect phenomenological test of naturalness depends on
the precision with which Higgs properties can be measured. At the LHC we
typically face uncertainties of ${\cal{O}}(10\%)$ due to dominant QCD
systematics \cite{Dittmaier:2011ti}.\footnote{These uncertainties may
  improve in the future, potentially reaching the ${\cal{O}}(1\%)$
  level in optimistic scenarios \cite{Incandela}.}  An indirect search
for natural physics is therefore best performed in the clean
phenomenological environment offered by a future lepton collider (LC),
although it is obviously not limited to such a machine.

Higgs associated production at a precision instrument like a LC
provides an extremely sensitive tool to analyze the Higgs. In
particular, when the collider is operated at $\sqrt{s}=250$~GeV,
associated production $e^+e^-\to hZ$, is the dominant Higgs production mode for $m_h\simeq
125$~GeV~\cite{Kilian:1995tr}.  The abundant production of Higgs
particles in a clean and fully reconstructible final state allows for
precise measurements of the Higgs couplings and properties
\cite{Peskin:2012we,ILCTRD}.  A particular strength of a precision associated production cross section
measurement lies in the fact that the $hZZ$ coupling can be determined
independent of Higgs decays, removing uncertainties in the total width
and other Higgs couplings. The program to reach a theoretically
precise understanding of $e^+e^-\to hZ$ in the Standard Model dates
back to the beginning of the LEP era
\cite{Fleischer:1982af,Jegerlehner:1983bf,Fleischer:1987zv,Denner:1992bc,Denner:2003iy}.
Recent analyses of the prospects for precision associated production cross section measurements
indicate that uncertainties as low as ${\cal{O}}(0.5 \%)$ can be
achieved \cite{Peskin:2012we,ILCTRD,Klute:2013cx,Blondel:2012ey}.\footnote{An accuracy of ${\cal{O}}(0.5 \%)$ would likely require reasonable assumptions on the total Higgs width; without such assumptions an accuracy of ${\cal{O}}(2.5 \%)$ or perhaps lower is possible.}

\begin{figure}[t!]
  \centering
  \includegraphics[width=0.47\textwidth]{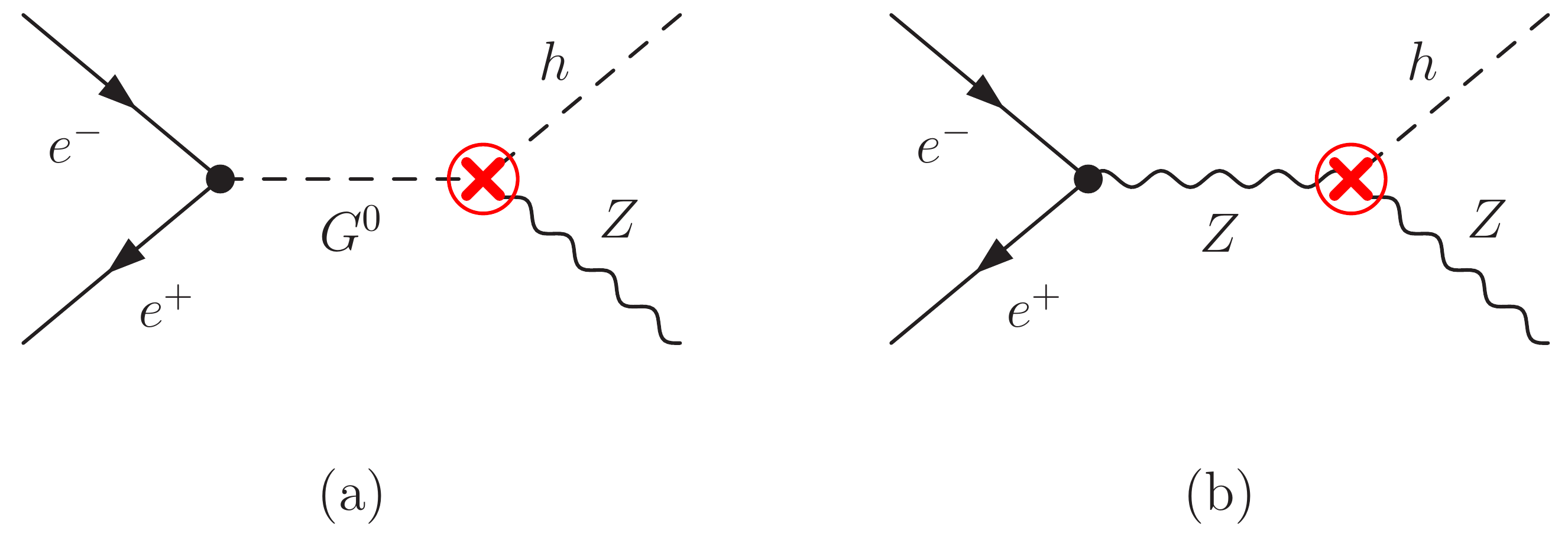}
  \caption{\label{fig:ct} Sample counterterm diagrams that depend on the
    Higgs self-energy.}\vskip 0.5cm
\end{figure}

BSM-modified NLO electroweak corrections to
associated production are typically larger than the projected
${\cal{O}}(0.5\%)$ uncertainty~\cite{Englert:2013tya}. Thus
Higgs boson coupling measurements can constrain natural new physics for
generic top partners {\it even when they are neutral under the SM
  gauge group.}  To see the relevant effects clearly, consider the
theory of \Eq{eq:L} when all scalar top partners, $\phi_i$, are gauge singlets. In the
limit $m_\phi \gg v$, we may integrate out the $\phi_i$ and express
their effects in terms of an effective Lagrangian below the scale
$m_\phi$ involving only Standard Model fields with appropriate
higher-dimensional operators. At one loop, integrating out the
$\phi_i$ leads to shifts in the wave-function renormalization and
potential of the Higgs doublet $H$ as well as operators of dimension
six and higher.  Most of these shifts and operators are irrelevant
from the perspective of low-energy physics, except for one
dimension-six operator in the effective Lagrangian:
\begin{equation} \label{eq:Leff}
  \mathcal{L}_{eff} = \mathcal{L}_{SM} + \frac{c_H}{m_\phi^2} \left( 
    \frac{1}{2} \partial_\mu |H|^2 \partial^\mu |H|^2 \right)+ \dots
\end{equation}
where the ellipses include additional higher-dimensional operators
that are irrelevant for our purposes. Matching to the full theory at
the scale $m_\phi$, we find $c_H(m_\phi) = n_\phi |\lambda_\phi|^2/96
\pi^2$. Although this operator may be exchanged for a linear
combination of other higher-dimensional operators using field
redefinitions or classical equations of motion, the physical effects
are unaltered. Below the scale of electroweak symmetry breaking, \Eq{eq:Leff}
leads to a shift in the wave-function renormalization of the physical
scalar $h$ as in \Eq{eq:Z}, with $\delta Z_h = 2 c_H v^2 /
m_\phi^2$. Canonically normalizing $h$ alters its coupling to vectors
and fermions, leading to a measurable correction to, e.g., the $hZ$
associated production cross-section
\begin{equation}
  \delta \sigma_{Zh} = - 2 c_H \frac{v^2}{m_\phi^2} = 
  - \frac{n_\phi |\lambda_\phi|^2}{48 \pi^2} \frac{v^2}{m_\phi^2}.
\end{equation}
where we have defined $\delta \sigma_{Zh}$ as the fractional change in the associated production 
cross section relative to the SM prediction, which by design vanishes for the SM alone.  Since $n_\phi |\lambda_\phi|^2$ is required to be large in order to cancel the top quadratic divergence, this effect may be observable in
precision measurements of $\sigma_{Zh}$ despite arising at one loop.

While this effective Lagrangian approach makes the physical effect
transparent, naturalness dictates that $m_\phi \sim v$, and threshold
corrections to \Eq{eq:Leff} may be large and a complete calculation is
required. 
In the on-shell renormalization scheme, the Higgs self-energy enters
through the counter-term part of the renormalized $e^+e^-\to hZ$
amplitude via the diagrams depicted in Fig.~\ref{fig:ct}.  Thus the $hG^0Z$
and $hZZ$ vertices receive corrections from the Higgs
wave-function renormalization.\footnote{See e.g.  Ref.~\cite{Denner:1991kt} for
a complete list of SM Feynman rules.}

\begin{figure}[]
  \centering
  \includegraphics[width=0.452\textwidth]{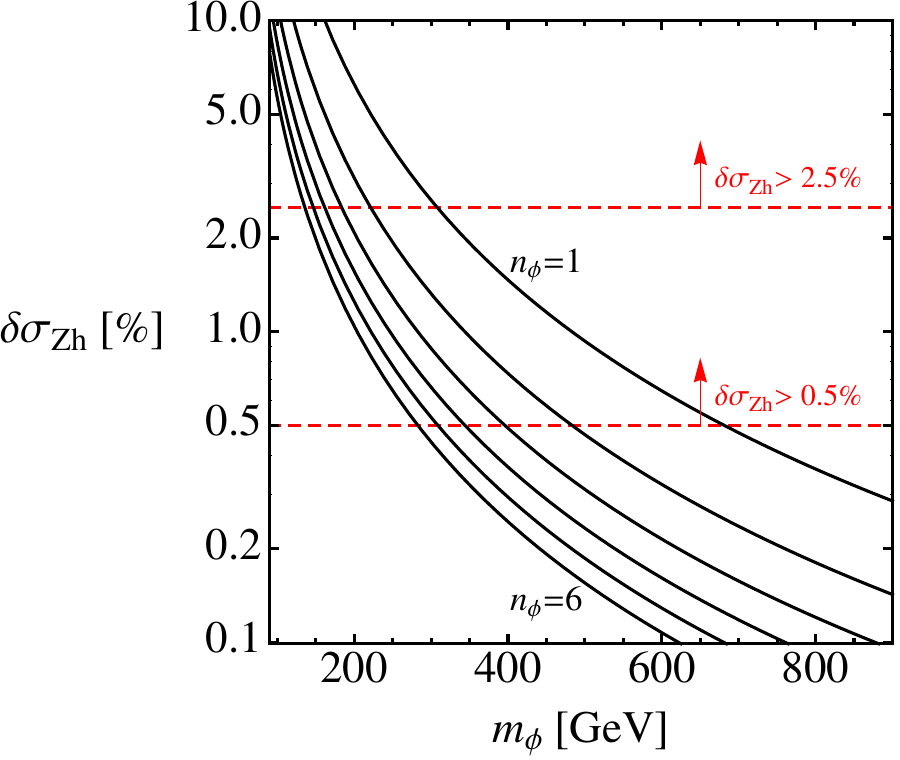}
  \caption{Scalar top-partner corrections to the Higgs associated
    production cross-section at a $250$ GeV linear collider as a
    function of the top-partner mass $m_\phi$ in the effective theory of naturalness of \Eq{eq:L}.  Corrections are shown
    for $n_\phi = 1,..,6$ top partners.  Estimates for the measurement precision of $2.5
    \%$ \cite{Peskin:2012we,ILCTRD} and $0.5\%$ \cite{Klute:2013cx}
    are also shown.  It is remarkable that with current precision
    estimates a large portion of model-independent parameter space for
    Higgs naturalness can be probed.  In particular, if one compares 
    with the tuning estimates of \Eq{eq:benchmarks}, this broadly corresponds to probing $10 \%$
    tuned regions for a single scalar top partner and close to $25 \%$ tuned regions
    for $n_\phi = 6$ scalar top partners as in SUSY.  Optimistically, if
    the precision could be improved to $\delta \sigma_{Zh} \sim 0.1
    \%$, then virtually all parameter space for
    generic natural scalar theories with up to $\sim10\%$ tunings could be
    probed.}
  \label{fig:corr}
\end{figure}

For scalar top partners the Higgs wave-function renormalization
arises at one loop through scalar trilinear couplings, which gauge
invariance relates to the quartic vertices, which are in turn
directly relevant for the cancellation of the quadratic
divergences in $\delta m_h^2$.

At one loop the effective theory of naturalness defined in \Eq{eq:L} leads to a
correction to the associated production cross-section of the form
\cite{Englert:2013tya}
\begin{eqnarray}
  \delta \sigma_{Zh} & = & n_\phi 
  \frac{|\lambda_\phi|^2 v^2}{8 \pi^2 m_h^2} \left(1+ F(\tau_{\phi}) \right) \\
  & = & \frac{9 \lambda_{t}^2 m_t^2}{2 \pi^2 n_\phi m_h^2} \left(1+ F(\tau_{\phi}) \right) 
\label{eq:approxloop}
\end{eqnarray}
where in the last line we have again imposed the cancellation of
quadratic divergences and $\tau_\phi = m_h^2/4 m_{\phi}^2$.  $F(\tau)$
is given by
\be
F(\tau)  =  \frac{1}{4 \sqrt{\tau (\tau-1)}} 
\log \left( \frac{  1-  2 \tau - 2 \sqrt{\tau (\tau-1)}}
  { 1- 2 \tau +2 \sqrt{\tau (\tau-1)} }\right).
\ee
\Eq{eq:approxloop} contains the full one-loop correction for gauge
singlet top-partners.  Additional corrections should also be included in the case where the top-partners
carry electroweak quantum numbers.  However,
these corrections have been calculated in \cite{Englert:2013tya} where
it was found that the one loop corrections are still dominated
by \Eq{eq:approxloop}.  This follows from the fact
that the square of the top Yukawa coupling is greater
than the square of electroweak couplings.  Thus \Eq{eq:approxloop}
applies equally well to generic scalar top-partners, irrespective of
gauge charges.

In \Fig{fig:corr} we show the extent to which the parameter space of
natural theories can be indirectly explored at a Linear Collider.  For
measurements of the associated production cross-section at the
estimated accuracy of $\mathcal{O} (0.5\%),$ natural theories tuned at
the $25\%$ level or greater can be probed, depending on the number of
degrees of freedom.  Optimistically, if the measurement accuracy were improved further
to $\mathcal{O} (0.1\%)$ then natural theories can be probed if they are tuned up to the $10\%$
level, even if they contain only
gauge singlets.  These results apply to the broad class of effective natural
theories described here, regardless of the top-partner gauge charges,
and hence contain SUSY theories as a subset.  They also apply to scenarios where top-partners
are difficult to directly discover or constrain due to their kinematic properties or quantum numbers.

If precision measurements of the Higgs associated production
cross-section at a linear collider show deviations from SM
expectations at the level of $\mathcal{O}(1\%)$ then this would
constitute strong indirect evidence for new physics in the Higgs
sector, and would be suggestive of a solution to the hierarchy
problem.  Alternatively, if no deviations are observed then such measurements could
put the compelling notion of electroweak naturalness under strain.

\acknowledgments We would like to thank Anson Hook, Markus Klute,
Dmytro Kovalskyi, Michael Peskin, Michael Rauch, Jesse Thaler, and Scott Thomas for
useful conversations. NC and MM acknowledge the hospitality of the
Weinberg Theory Group at the University of Texas, Austin, where this
work was initiated.

\bibliography{NaturalLCref}

\begin{thebibliography}{31}
\expandafter\ifx\csname natexlab\endcsname\relax\def\natexlab#1{#1}\fi
\expandafter\ifx\csname bibnamefont\endcsname\relax
  \def\bibnamefont#1{#1}\fi
\expandafter\ifx\csname bibfnamefont\endcsname\relax
  \def\bibfnamefont#1{#1}\fi
\expandafter\ifx\csname citenamefont\endcsname\relax
  \def\citenamefont#1{#1}\fi
\expandafter\ifx\csname url\endcsname\relax
  \def\url#1{\texttt{#1}}\fi
\expandafter\ifx\csname urlprefix\endcsname\relax\def\urlprefix{URL }\fi
\providecommand{\bibinfo}[2]{#2}
\providecommand{\eprint}[2][]{\url{#2}}

\bibitem[{ATL(2013)}]{ATLAS-CONF-2013-034}
\bibinfo{type}{Tech. Rep.} \bibinfo{number}{ATLAS-CONF-2013-034},
  \bibinfo{institution}{CERN}, \bibinfo{address}{Geneva}
  (\bibinfo{year}{2013}).

\bibitem[{CMS(2013)}]{CMS-PAS-HIG-13-005}
\bibinfo{type}{Tech. Rep.} \bibinfo{number}{CMS-PAS-HIG-13-005},
  \bibinfo{institution}{CERN}, \bibinfo{address}{Geneva}
  (\bibinfo{year}{2013}).

\bibitem[{\citenamefont{Chacko et~al.}(2006)\citenamefont{Chacko, Goh, and
  Harnik}}]{Chacko:2005pe}
\bibinfo{author}{\bibfnamefont{Z.}~\bibnamefont{Chacko}},
  \bibinfo{author}{\bibfnamefont{H.-S.} \bibnamefont{Goh}}, \bibnamefont{and}
  \bibinfo{author}{\bibfnamefont{R.}~\bibnamefont{Harnik}},
  \bibinfo{journal}{Phys.Rev.Lett.} \textbf{\bibinfo{volume}{96}},
  \bibinfo{pages}{231802} (\bibinfo{year}{2006}), \eprint{hep-ph/0506256}.

\bibitem[{\citenamefont{Burdman et~al.}(2007)\citenamefont{Burdman, Chacko,
  Goh, and Harnik}}]{Burdman:2006tz}
\bibinfo{author}{\bibfnamefont{G.}~\bibnamefont{Burdman}},
  \bibinfo{author}{\bibfnamefont{Z.}~\bibnamefont{Chacko}},
  \bibinfo{author}{\bibfnamefont{H.-S.} \bibnamefont{Goh}}, \bibnamefont{and}
  \bibinfo{author}{\bibfnamefont{R.}~\bibnamefont{Harnik}},
  \bibinfo{journal}{JHEP} \textbf{\bibinfo{volume}{0702}}, \bibinfo{pages}{009}
  (\bibinfo{year}{2007}), \eprint{hep-ph/0609152}.

\bibitem[{\citenamefont{Arvanitaki and Villadoro}(2012)}]{Arvanitaki:2011ck}
\bibinfo{author}{\bibfnamefont{A.}~\bibnamefont{Arvanitaki}} \bibnamefont{and}
  \bibinfo{author}{\bibfnamefont{G.}~\bibnamefont{Villadoro}},
  \bibinfo{journal}{JHEP} \textbf{\bibinfo{volume}{1202}}, \bibinfo{pages}{144}
  (\bibinfo{year}{2012}), \eprint{1112.4835}.

\bibitem[{\citenamefont{Blum et~al.}(2013)\citenamefont{Blum, D'Agnolo, and
  Fan}}]{Blum:2012ii}
\bibinfo{author}{\bibfnamefont{K.}~\bibnamefont{Blum}},
  \bibinfo{author}{\bibfnamefont{R.~T.} \bibnamefont{D'Agnolo}},
  \bibnamefont{and} \bibinfo{author}{\bibfnamefont{J.}~\bibnamefont{Fan}},
  \bibinfo{journal}{JHEP} \textbf{\bibinfo{volume}{1301}}, \bibinfo{pages}{057}
  (\bibinfo{year}{2013}), \eprint{1206.5303}.

\bibitem[{\citenamefont{Montull and Riva}(2012)}]{Montull:2012ik}
\bibinfo{author}{\bibfnamefont{M.}~\bibnamefont{Montull}} \bibnamefont{and}
  \bibinfo{author}{\bibfnamefont{F.}~\bibnamefont{Riva}},
  \bibinfo{journal}{JHEP} \textbf{\bibinfo{volume}{1211}}, \bibinfo{pages}{018}
  (\bibinfo{year}{2012}), \eprint{1207.1716}.

\bibitem[{\citenamefont{Peskin and Takeuchi}(1990)}]{Peskin:1990zt}
\bibinfo{author}{\bibfnamefont{M.~E.} \bibnamefont{Peskin}} \bibnamefont{and}
  \bibinfo{author}{\bibfnamefont{T.}~\bibnamefont{Takeuchi}},
  \bibinfo{journal}{Phys.Rev.Lett.} \textbf{\bibinfo{volume}{65}},
  \bibinfo{pages}{964} (\bibinfo{year}{1990}).

\bibitem[{\citenamefont{Peskin and Takeuchi}(1992)}]{Peskin:1991sw}
\bibinfo{author}{\bibfnamefont{M.~E.} \bibnamefont{Peskin}} \bibnamefont{and}
  \bibinfo{author}{\bibfnamefont{T.}~\bibnamefont{Takeuchi}},
  \bibinfo{journal}{Phys.Rev.} \textbf{\bibinfo{volume}{D46}},
  \bibinfo{pages}{381} (\bibinfo{year}{1992}).

\bibitem[{\citenamefont{Altarelli and Barbieri}(1991)}]{Altarelli:1990zd}
\bibinfo{author}{\bibfnamefont{G.}~\bibnamefont{Altarelli}} \bibnamefont{and}
  \bibinfo{author}{\bibfnamefont{R.}~\bibnamefont{Barbieri}},
  \bibinfo{journal}{Phys.Lett.} \textbf{\bibinfo{volume}{B253}},
  \bibinfo{pages}{161} (\bibinfo{year}{1991}).

\bibitem[{\citenamefont{Altarelli et~al.}(1992)\citenamefont{Altarelli,
  Barbieri, and Jadach}}]{Altarelli:1991fk}
\bibinfo{author}{\bibfnamefont{G.}~\bibnamefont{Altarelli}},
  \bibinfo{author}{\bibfnamefont{R.}~\bibnamefont{Barbieri}}, \bibnamefont{and}
  \bibinfo{author}{\bibfnamefont{S.}~\bibnamefont{Jadach}},
  \bibinfo{journal}{Nucl.Phys.} \textbf{\bibinfo{volume}{B369}},
  \bibinfo{pages}{3} (\bibinfo{year}{1992}).

\bibitem[{\citenamefont{Kennedy and Lynn}(1989)}]{Kennedy:1988sn}
\bibinfo{author}{\bibfnamefont{D.}~\bibnamefont{Kennedy}} \bibnamefont{and}
  \bibinfo{author}{\bibfnamefont{B.}~\bibnamefont{Lynn}},
  \bibinfo{journal}{Nucl.Phys.} \textbf{\bibinfo{volume}{B322}},
  \bibinfo{pages}{1} (\bibinfo{year}{1989}).

\bibitem[{\citenamefont{Kennedy et~al.}(1989)\citenamefont{Kennedy, Lynn, Im,
  and Stuart}}]{Kennedy:1988rt}
\bibinfo{author}{\bibfnamefont{D.}~\bibnamefont{Kennedy}},
  \bibinfo{author}{\bibfnamefont{B.}~\bibnamefont{Lynn}},
  \bibinfo{author}{\bibfnamefont{C.}~\bibnamefont{Im}}, \bibnamefont{and}
  \bibinfo{author}{\bibfnamefont{R.}~\bibnamefont{Stuart}},
  \bibinfo{journal}{Nucl.Phys.} \textbf{\bibinfo{volume}{B321}},
  \bibinfo{pages}{83} (\bibinfo{year}{1989}).

\bibitem[{\citenamefont{Barbieri et~al.}(2004)\citenamefont{Barbieri, Pomarol,
  Rattazzi, and Strumia}}]{Barbieri:2004qk}
\bibinfo{author}{\bibfnamefont{R.}~\bibnamefont{Barbieri}},
  \bibinfo{author}{\bibfnamefont{A.}~\bibnamefont{Pomarol}},
  \bibinfo{author}{\bibfnamefont{R.}~\bibnamefont{Rattazzi}}, \bibnamefont{and}
  \bibinfo{author}{\bibfnamefont{A.}~\bibnamefont{Strumia}},
  \bibinfo{journal}{Nucl.Phys.} \textbf{\bibinfo{volume}{B703}},
  \bibinfo{pages}{127} (\bibinfo{year}{2004}), \eprint{hep-ph/0405040}.

\bibitem[{\citenamefont{Englert and McCullough}(2013)}]{Englert:2013tya}
\bibinfo{author}{\bibfnamefont{C.}~\bibnamefont{Englert}} \bibnamefont{and}
  \bibinfo{author}{\bibfnamefont{M.}~\bibnamefont{McCullough}}
  (\bibinfo{year}{2013}), \eprint{1303.1526}.

\bibitem[{\citenamefont{Kitano and Nomura}(2006)}]{Kitano:2006gv}
\bibinfo{author}{\bibfnamefont{R.}~\bibnamefont{Kitano}} \bibnamefont{and}
  \bibinfo{author}{\bibfnamefont{Y.}~\bibnamefont{Nomura}},
  \bibinfo{journal}{Phys.Rev.} \textbf{\bibinfo{volume}{D73}},
  \bibinfo{pages}{095004} (\bibinfo{year}{2006}), \eprint{hep-ph/0602096}.

\bibitem[{\citenamefont{Giardino et~al.}(2013)\citenamefont{Giardino, Kannike,
  Masina, Raidal, and Strumia}}]{Giardino:2013bma}
\bibinfo{author}{\bibfnamefont{P.~P.} \bibnamefont{Giardino}},
  \bibinfo{author}{\bibfnamefont{K.}~\bibnamefont{Kannike}},
  \bibinfo{author}{\bibfnamefont{I.}~\bibnamefont{Masina}},
  \bibinfo{author}{\bibfnamefont{M.}~\bibnamefont{Raidal}}, \bibnamefont{and}
  \bibinfo{author}{\bibfnamefont{A.}~\bibnamefont{Strumia}}
  (\bibinfo{year}{2013}), \eprint{1303.3570}.

\bibitem[{\citenamefont{Djouadi et~al.}(2012)\citenamefont{Djouadi, Lebedev,
  Mambrini, and Quevillon}}]{Djouadi:2011aa}
\bibinfo{author}{\bibfnamefont{A.}~\bibnamefont{Djouadi}},
  \bibinfo{author}{\bibfnamefont{O.}~\bibnamefont{Lebedev}},
  \bibinfo{author}{\bibfnamefont{Y.}~\bibnamefont{Mambrini}}, \bibnamefont{and}
  \bibinfo{author}{\bibfnamefont{J.}~\bibnamefont{Quevillon}},
  \bibinfo{journal}{Phys.Lett.} \textbf{\bibinfo{volume}{B709}},
  \bibinfo{pages}{65} (\bibinfo{year}{2012}), \eprint{1112.3299}.

\bibitem[{\citenamefont{Dittmaier et~al.}(2011)}]{Dittmaier:2011ti}
\bibinfo{author}{\bibfnamefont{S.}~\bibnamefont{Dittmaier}}
  \bibnamefont{et~al.} (\bibinfo{collaboration}{LHC Higgs Cross Section Working
  Group}) (\bibinfo{year}{2011}), \eprint{1101.0593}.

\bibitem[{\citenamefont{Incandela}(2013)}]{Incandela}
\bibinfo{author}{\bibfnamefont{J.}~\bibnamefont{Incandela}},
  \bibinfo{journal}{{Talk at the ``Higgs After Discovery Workshop'',
  Princeton.}}  (\bibinfo{year}{2013}).

\bibitem[{\citenamefont{Kilian et~al.}(1996)\citenamefont{Kilian, Kramer, and
  Zerwas}}]{Kilian:1995tr}
\bibinfo{author}{\bibfnamefont{W.}~\bibnamefont{Kilian}},
  \bibinfo{author}{\bibfnamefont{M.}~\bibnamefont{Kramer}}, \bibnamefont{and}
  \bibinfo{author}{\bibfnamefont{P.}~\bibnamefont{Zerwas}},
  \bibinfo{journal}{Phys.Lett.} \textbf{\bibinfo{volume}{B373}},
  \bibinfo{pages}{135} (\bibinfo{year}{1996}), \eprint{hep-ph/9512355}.

\bibitem[{\citenamefont{Peskin}(2012)}]{Peskin:2012we}
\bibinfo{author}{\bibfnamefont{M.~E.} \bibnamefont{Peskin}}
  (\bibinfo{year}{2012}), \eprint{1207.2516}.

\bibitem[{\citenamefont{{Physics at the International Linear
  Collider}}(2013)}]{ILCTRD}
\bibinfo{author}{\bibnamefont{{Physics at the International Linear Collider}}},
  \bibinfo{journal}{http://ific.uv.es/~fuster/DBD-Chapters/}
  (\bibinfo{year}{2013}).

\bibitem[{\citenamefont{Fleischer and Jegerlehner}(1983)}]{Fleischer:1982af}
\bibinfo{author}{\bibfnamefont{J.}~\bibnamefont{Fleischer}} \bibnamefont{and}
  \bibinfo{author}{\bibfnamefont{F.}~\bibnamefont{Jegerlehner}},
  \bibinfo{journal}{Nucl.Phys.} \textbf{\bibinfo{volume}{B216}},
  \bibinfo{pages}{469} (\bibinfo{year}{1983}).

\bibitem[{\citenamefont{Jegerlehner}(1983)}]{Jegerlehner:1983bf}
\bibinfo{author}{\bibfnamefont{F.}~\bibnamefont{Jegerlehner}}
  (\bibinfo{year}{1983}).

\bibitem[{\citenamefont{Fleischer and Jegerlehner}(1987)}]{Fleischer:1987zv}
\bibinfo{author}{\bibfnamefont{J.}~\bibnamefont{Fleischer}} \bibnamefont{and}
  \bibinfo{author}{\bibfnamefont{F.}~\bibnamefont{Jegerlehner}}
  (\bibinfo{year}{1987}).

\bibitem[{\citenamefont{Denner et~al.}(1992)\citenamefont{Denner, Kublbeck,
  Mertig, and Bohm}}]{Denner:1992bc}
\bibinfo{author}{\bibfnamefont{A.}~\bibnamefont{Denner}},
  \bibinfo{author}{\bibfnamefont{J.}~\bibnamefont{Kublbeck}},
  \bibinfo{author}{\bibfnamefont{R.}~\bibnamefont{Mertig}}, \bibnamefont{and}
  \bibinfo{author}{\bibfnamefont{M.}~\bibnamefont{Bohm}},
  \bibinfo{journal}{Z.Phys.} \textbf{\bibinfo{volume}{C56}},
  \bibinfo{pages}{261} (\bibinfo{year}{1992}).

\bibitem[{\citenamefont{Denner et~al.}(2003)\citenamefont{Denner, Dittmaier,
  Roth, and Weber}}]{Denner:2003iy}
\bibinfo{author}{\bibfnamefont{A.}~\bibnamefont{Denner}},
  \bibinfo{author}{\bibfnamefont{S.}~\bibnamefont{Dittmaier}},
  \bibinfo{author}{\bibfnamefont{M.}~\bibnamefont{Roth}}, \bibnamefont{and}
  \bibinfo{author}{\bibfnamefont{M.}~\bibnamefont{Weber}},
  \bibinfo{journal}{Nucl.Phys.} \textbf{\bibinfo{volume}{B660}},
  \bibinfo{pages}{289} (\bibinfo{year}{2003}), \eprint{hep-ph/0302198}.

\bibitem[{\citenamefont{Klute et~al.}(2013)\citenamefont{Klute, Lafaye, Plehn,
  Rauch, and Zerwas}}]{Klute:2013cx}
\bibinfo{author}{\bibfnamefont{M.}~\bibnamefont{Klute}},
  \bibinfo{author}{\bibfnamefont{R.}~\bibnamefont{Lafaye}},
  \bibinfo{author}{\bibfnamefont{T.}~\bibnamefont{Plehn}},
  \bibinfo{author}{\bibfnamefont{M.}~\bibnamefont{Rauch}}, \bibnamefont{and}
  \bibinfo{author}{\bibfnamefont{D.}~\bibnamefont{Zerwas}},
  \bibinfo{journal}{Europhys.Lett.} \textbf{\bibinfo{volume}{101}},
  \bibinfo{pages}{51001} (\bibinfo{year}{2013}), \eprint{1301.1322}.

\bibitem[{\citenamefont{Blondel et~al.}(2012)\citenamefont{Blondel, Koratzinos,
  Assmann, Butterworth, Janot et~al.}}]{Blondel:2012ey}
\bibinfo{author}{\bibfnamefont{A.}~\bibnamefont{Blondel}},
  \bibinfo{author}{\bibfnamefont{M.}~\bibnamefont{Koratzinos}},
  \bibinfo{author}{\bibfnamefont{R.}~\bibnamefont{Assmann}},
  \bibinfo{author}{\bibfnamefont{A.}~\bibnamefont{Butterworth}},
  \bibinfo{author}{\bibfnamefont{P.}~\bibnamefont{Janot}}, \bibnamefont{et~al.}
  (\bibinfo{year}{2012}), \eprint{1208.0504}.

\bibitem[{\citenamefont{Denner}(1993)}]{Denner:1991kt}
\bibinfo{author}{\bibfnamefont{A.}~\bibnamefont{Denner}},
  \bibinfo{journal}{Fortsch.Phys.} \textbf{\bibinfo{volume}{41}},
  \bibinfo{pages}{307} (\bibinfo{year}{1993}), \eprint{0709.1075}.

\end{thebibliography}

\end{document}